\documentclass{article}
\usepackage{spconf,amsmath,graphicx}
\usepackage{amsfonts}
\usepackage{multirow}
\usepackage{microtype}
\usepackage{cite}
\usepackage{booktabs}
\usepackage[dvipsnames]{xcolor}
\def\v#1{\mathbf{#1}}
\def\l#1{\mathcal{#1}}
\DeclareMathOperator*{\argmax}{arg\,max}

\newcommand{\kledit}[1]{{#1}}
\newcommand{\bsedit}[1]{{#1}}
\newcommand{\ssedit}[1]{{#1}}

\newcommand{\klnewedit}[1]{\textcolor{blue}{#1}}
\newcommand{\bsnewedit}[1]{\textcolor{magenta}{#1}}

\newcommand{\klcomment}[1]{\textcolor{red}{#1 (KL)}}
\newcommand{\bscomment}[1]{\textcolor{teal}{#1 (BS)}}
\newcommand{\sscomment}[1]{\textcolor{RedOrange}{#1 (SS)}}

\renewcommand{\klcomment}[1]{{}}
\renewcommand{\bscomment}[1]{{}}
\renewcommand{\sscomment}[1]{{}}

\renewcommand{\klnewedit}[1]{{#1}}
\renewcommand{\bsnewedit}[1]{{#1}}

\title{Whole-Word Segmental Speech Recognition with Acoustic Word Embeddings}
\name{Bowen Shi, Shane Settle, Karen Livescu}
\address{TTI-Chicago, USA\\\small\texttt{\{bshi,settle.shane,klivescu\}@ttic.edu}}
\begin{document}
\maketitle

\begin{abstract}
Segmental models are 
\kledit{sequence prediction models in which}
scores of hypotheses are based on
entire variable-length segments of frames.
We consider segmental models for whole-word (``acoustic-to-word'') speech recognition, with the \klnewedit{feature vectors} %for a segment 
defined using \klnewedit{vector embeddings of segments}.
\kledit{Such} models are computationally challenging as the number of paths is proportional to the vocabulary size, which can be orders of magnitude larger than when using \kledit{subword} units like phones.
We describe an efficient approach for end-to-end whole-word segmental models, 
\kledit{with} forward-backward and Viterbi decoding 
performed
on a GPU and a simple 
\kledit{segment scoring} function 
\kledit{that} reduces space complexity. 
In addition, we investigate the use of \kledit{pre-training via jointly trained acoustic word embeddings (AWEs) and acoustically grounded word embeddings (AGWEs) of written word labels.} 
\ssedit{We find that word error rate can be reduced by a large margin by pre-training the acoustic} \klnewedit{segment} \kledit{representation with AWEs}, \kledit{and additional (smaller) gains can be obtained by pre-training the word prediction layer with AGWEs.  Our final models improve over 
\bsnewedit{prior} A2W \kledit{models.}}
\end{abstract}
\begin{keywords}
speech recognition, segmental model, acoustic-to-word, acoustic word embeddings, pre-training
\end{keywords}

\vspace{-0.4cm}
\section{Introduction}
\vspace{-0.2cm}
Acoustic-to-word (A2W) models for speech recognition map input acoustic frames directly to words. Unlike conventional subword-based automatic speech recognition (ASR) systems, A2W models do not require an external 
\kledit{lexicon}, 
thus simplifying training and decoding.
Recent work has shown that A2W models can achieve performance competitive with state-of-the-art subword-based systems either with large amounts of training data \cite{soltau_a2w} or with careful training techniques \cite{audhkhasi2017_a2w,audhkhasi2018_a2w,yu2018_amt,settle2019_a2w}.

Most work on A2W models~\cite{soltau_a2w,audhkhasi2017_a2w,audhkhasi2018_a2w,li2018advancing,yu2018_amt,settle2019_a2w,gaur2019acoustic} is based on connectionist temporal classification (CTC)~\cite{graves2006connectionist}, where the word sequence probability is defined as the product of frame-level 
probabilities.
In such approaches there is no explicit modeling of \klnewedit{segments} of frames corresponding to words. \kledit{There has also been recent work on encoder-decoder A2W models, which can focus on \klnewedit{"soft segments"}
via an attention mechanism~\cite{collobert2019wordlevel,palaskar2018_s2s}.} 

\begin{figure}[t]
  \center
  \includegraphics[width=0.95\linewidth]{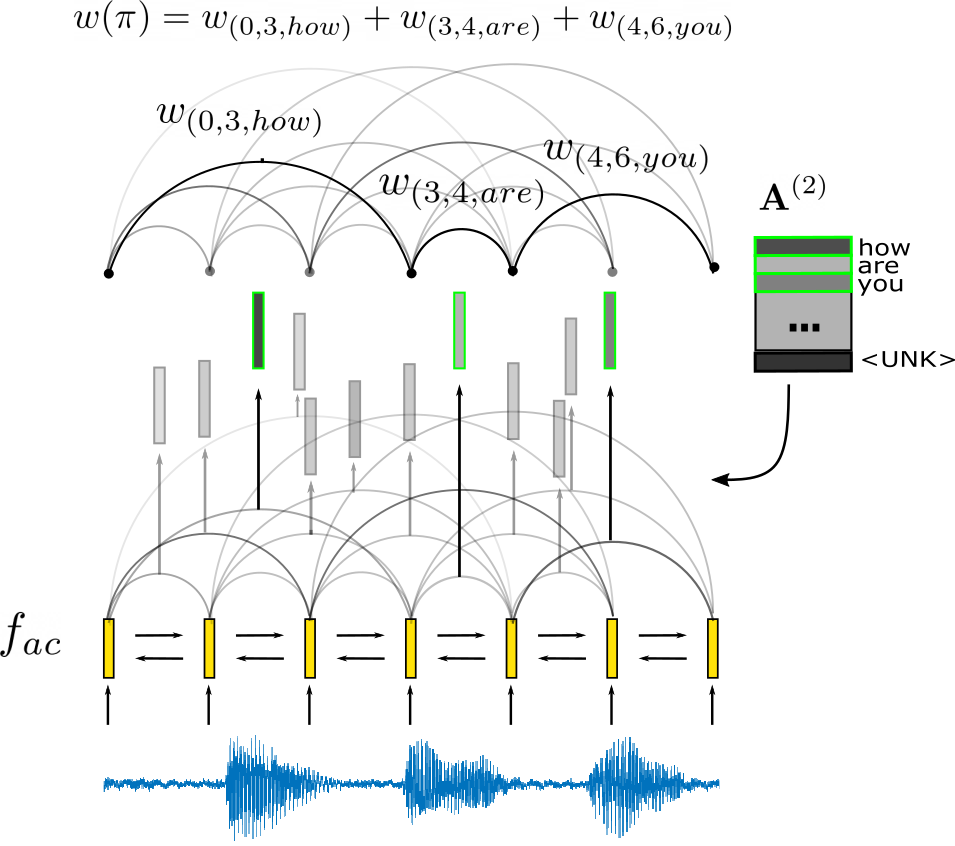}
  \vspace{-0.3cm}
  \caption{\vspace{-0.0cm}\label{fig:model}Whole-word segmental model for speech recognition. Note: boundary frames are not shared. 
  \vspace{-0.7cm}
  }
\end{figure}

In this paper we \kledit{propose} an approach \kledit{using} whole-word segmental models, where the sequence probability is computed 
\bsedit{based on {\it segment} scores instead of {\it frame} probabilities.}
  \kledit{Segmental models have a long history in speech recognition research, but they have been used primarily for phonetic recognition
or as \kledit{phone-level acoustic models}
~\cite{ostendorf1996hmm,glass2003probabilistic,zweig2009segmental,zweig2012_seg,he2012_sc,hamid2013_dsnn,tang2014_losses,lu2016_srnn}}. There has also been 
work on whole-word segmental models for second-pass rescoring~\cite{zweig2009segmental,maas2012word,bengio2014word}, but to our knowledge our approach is the first to address end-to-end A2W segmental models.

\kledit{The key ingredient in our approach is to define the segment scores in terms of dot products between} 
\klnewedit{vector embeddings of acoustic segments and a weight layer of written word embeddings.} 
This form of the model allows for (1) efficient re-use of feature functions and therefore reduced memory cost 
and (2) \klnewedit{initialization of the acoustic and written embeddings using pre-trained acoustic word embeddings (AWEs) and acoustically grounded word embeddings (AGWEs), } 
following the successful use of such pre-training in prior work on speech recognition~\cite{settle2019_a2w} and search~\cite{settle2017query,audhkhasi2017end}. 
\kledit{We also obtain speed-ups via GPU implementations of the forward-backward and Viterbi algorithms.}
\kledit{We find that} pre-trained AWEs provide \kledit{large} gains, and result in segmental models that outperform the best prior A2W models on conversational telephone speech recognition.

\section{Segmental Model Formulation}
\vspace{-.2cm}
Segmental models compute \kledit{the score of a hypothesized label sequence as a combination of scores of multi-frame segments of speech in the sequence, rather than using individual frame scores (see Figure~\ref{fig:model}).} 
Let $\v X=\{\v  x_1, \v x_2, ..., \v x_T\}$  be a sequence of input acoustic frames and $\v L=\{l_1, l_2, ..., l_K\}$
be the output label sequence. A segmentation $\pi$ with respect to ${\bf X}$ and ${\bf L}$ is defined as a sequence of tuples 
\bsedit{$\{(t_1, s_1, l_1), (t_2, s_2, l_2), ..., (t_K, s_K, l_K)\}$.}
 Each \kledit{tuple} 
defines a segment $e_k$ \kledit{\bsedit{consisting of a start timestep\footnote{Frame $x_t$ is the acoustic signal between timesteps $t-1$ and $t$.} $t_k$, an end timestep $t_k+s_k$, and a label $l_k$, such that $t_1=0, t_K+s_K=T, t_k+s_k=t_{k+1},$ and $s_k>0$ for all $1\leq k\leq K$.}}
A segmental model assigns a \bsedit{score} $w_{t,s,v}$ to each segment $(t,s,v)$. The \bsedit{score} of a segmentation is then defined as 
\bsedit{$w(\pi)= \sum_{(t,s,v)\in\pi}w_{t,s,v}$.}
\vspace{-0.3cm}
\subsection{\kledit{Segment Score Functions}}
\vspace{-0.1cm}
As in other \klnewedit{recent} sequence models,
\ssedit{the input acoustic frames 
are first passed through a neural network
and encoded into frame features $\v H = \text{Enc}(\v X) \in \mathbb{R}^{T\times F}$\bsedit{, where $F$ denotes the feature \klnewedit{dimensionality}.} 
In segmental \klnewedit{models}, however, these frame features are then used to produce segment scores $\v W \in \mathbb{R}^{T \times S \times V}$, where $S$ and $V$
denote the maximum segment size and vocabulary size, 
respectively, and $w_{t,s,v}$ is the \kledit{score} of segment \bsedit{$(t, s, v)$}.
}
\ssedit{
Our approach defines segment scores ${\bf W}$ 
\kledit{in terms of} dot products between learned representations of variable-length segments and word labels:
  \vspace{-0.2cm}
\begin{flalign}
  \label{eq:feat_func}
w_{t,s,v} &= {\v a^{(2)}_v}^T f_{ac}(\v H_{t:t+s}) + b^{(2)}_v
\end{flalign}
where 
\klnewedit{$f_{ac}$ is an acoustic segment embedding} function mapping segments $\v H_{t:t+s} \in \mathbb{R}^{s \times F}$ to fixed-dimensional embeddings 
\klnewedit{$f_{ac}(\v H_{t:t+s}) \in \mathbb{R}^D$, $\v a^{(2)}_v$} is a row from the matrix $\v A^{(2)} \in \mathbb{R}^{V \times D}$ composed of 
\klnewedit{embeddings} for \kledit{all words} $v$ in the vocabulary, \bsedit{and $b_v$ is the bias on word $v$, which can be interpreted as a log-unigram probability}. 
\klnewedit{We define the acoustic segment embedding function} as follows: \vspace{-0.1cm}
\begin{flalign}
  f_{ac}(\v H_{t:t+s}) &= \text{ReLU}(\v A^{(1)} G(\v H_{t:t+s}) + \v b^{(1)})
\label{eq:awe_func}
\end{flalign}
\vspace{-0.1cm}
where $G$ is a pooling function \ssedit{chosen between:} 
\vspace{-0.0cm}
\begin{flalign}
&\hspace{-0.2cm}G(\v H_{t:t+s}) =[\v h_t; \v h_{t+s}] \label{eq:concat}\\
&\hspace{-0.2cm}G(\v H_{t:t+s}) = \frac{1}{s}\textstyle\sum_{i=1}^{s}\v h_{t+i} \label{eq:avg}\\
&\hspace{-0.2cm}G(\v H_{t:t+s}) =\frac{1}{s}\textstyle\sum_{i=1}^{s}{\text{Softmax}(\v g^{T}\v H_{t:t+s})}_i\v h_{t+i}\label{eq:attn}
\end{flalign}
\vspace{-0.1cm}
where (\ref{eq:concat}) is concatenation, (\ref{eq:avg}) is \klnewedit{mean} pooling, and (\ref{eq:attn}) is attention pooling (with learnable parameter {\bf g}).
Equation~\ref{eq:feat_func} allows 
feature sharing, which helps 
limit the memory 
\klnewedit{needed to compute} segment features to $O(TSD)$  and simplifies scoring to matrix multiplication, i.e. 
\klnewedit{$\v W_{t,s}=\v A^{(2)} f_{ac}(\v H_{t:t+s})+\v b^{(2)}$.}}

\kledit{Recent work on segmental models has largely used two types of segment score functions:  (1) frame classifier-based~\cite{hamid2013_dsnn,tang2015_cascade,he2012_sc,tang2017_e2e} and (2) segmental recurrent neural network (SRNN)~\cite{lu2016_srnn,kong2016_srnn,tang2017_e2e}. Frame classifier-based score functions use a mapping from input acoustic frames $\v X$ to frame \bsedit{log-probability} vectors $\v P$, which are then pooled (via mean, sampling, etc.) to get the segment \bsedit{score} $w_{t,s,v}$.
This method  
introduces a multiplicative memory dependence on $V$, \kledit{which is} a factor $V/D$ 
increase in memory overhead over our approach.   
In our case $V$ is the number of words in the vocabulary, 
\bsnewedit{which is typically $\sim 10$ times larger than $D$ and makes this approach \klnewedit{extremely (sometimes prohibitively)} expensive.} 
\klnewedit{SRNNs} 
compute the \kledit{score} $w_{t, s, v} = \phi^T f_{\theta}([\v h_t; \v h_{t+s}; \v a^{(2)}_v])$,
where $f_{\theta}$ is a learned feature function, $\v a^{(2)}_v$ is an embedding of word $v$, and $[\v u;\v v]$ denotes concatenation of $\v u,\v v$. This method introduces an $O(TSDV)$ memory overhead, which can again quickly make it infeasible for \kledit{large-vocabulary recognition.} 
}

\kledit{
In addition to computational savings, our formulation of segment scores 
\klnewedit{in terms of products of acoustic embeddings and written word embeddings} also has the advantage that these two factors can be pre-trained using methods 
\kledit{from} prior work~\cite{he+etal_iclr2017,settle2019_a2w} (see Section~\ref{ssec:agwe}).
}

\vspace{-0.3cm}
\subsection{Training}
\kledit{Segmental models can be trained in a variety of ways~\cite{tang2017_e2e}.  One way, which we adopt here, is to interpret %segmental models 
them as probabilistic models and optimize the marginal log loss under that model, which is equivalent to viewing our models as segmental conditional random fields~\cite{zweig2011speech}.  Under this view, the model assigns probabilities to paths, conditioned on the input acoustic sequence, by normalizing the path score.
Letting $\v U:=\exp(\v W)$, we define
$p(\pi):=\frac{u(\pi)}{\sum_{\pi\in\mathcal{P}_{0:T}}u(\pi)}$ as the probability of the segmentation,} where $u(\pi) =\prod_{(t,s,v)\in\pi} u_{t,s,v}$ \kledit{and} $\mathcal{P}_{0:T}$ denotes all segmentations of $\v X_{1:T}$. 
\klnewedit{We define} the loss for a given word sequence $\v L$ and input $\v X$ as the marginal log loss,
by marginalizing \kledit{over} all possible \kledit{segmentations:}
\vspace{-0.12cm}
\begin{equation}
  \begin{split}
    \label{eq:log_loss}
    \l L(\v L, \v X) &= -\log\displaystyle\sum_{\substack{\pi\in\mathcal{P}_{0:T} \text{,}\\ \mathcal{B}(\pi)=\v L}}u(\pi) +\log{\displaystyle\sum_{\pi\in\mathcal{P}_{0:T}}u(\pi)} \\
  \end{split}
\end{equation}
%\vspace{-0.12cm}
where \bsedit{$\mathcal{B}(\pi)$ maps $\pi=\{(t_k,s_k,l_k)\}_{1\leq k\leq |\pi|}$ to its label sequence $\{l_i\}_{1\leq k\leq |\pi|}$.}
The summations 
can be efficiently computed with dynamic programming:
  \vspace{-0.15cm}
\begin{equation}
  \vspace{-0.15cm}
  \label{eq:alpha_nom_denom}
    \begin{split}
\alpha^{(d)}_t &:= \displaystyle\sum_{\pi\in\l P_{0:t}}u(\pi) = \displaystyle\sum_{ s=1}^S\displaystyle\sum_{v=1}^{V}u_{t-s,s,v}\alpha^{(d)}_{t-s} \\[-5pt]
\alpha^{(n)}_{t,y} &:= \hspace{-10pt}\displaystyle\sum_{\substack{\pi\in\l P_{0:t} \\ \l B(\pi_{1:y})=\v L_{1:y}}}\hspace{-10pt}u(\pi) = \displaystyle\sum_{s=1}^S u_{t-s,s, l_y}\alpha^{(n)}_{t-s, y-1}\\
    \end{split}
\end{equation}
With $\alpha^{(d)}$ and $\alpha^{(n)}$ computed, the loss value follows directly from 
$\l L(\v L, \v X)=-\log\alpha^{(n)}_{T,|\v L|}+\log{\alpha^{(d)}_T}$.
The last summations in Equations~\ref{eq:alpha_nom_denom} can be efficiently implemented on a GPU. In addition, 
$\alpha_{1:T,y}^{(n)}$ can be computed in parallel given $\alpha_{1:T,y-1}^{(n)}$ \ssedit{such that} %. Thus,
the overall time complexity\footnote{Number of times $a+b$ is called} of computing the loss is $O(T\log(SV)+|\v L|\log(S))$. 

To train with gradient descent, we need to differentiate $\l L(\v L, \v X)$ with respect to $\v X$, \kledit{which can in principle be done with auto-differentiation toolkits (e.g.~PyTorch~\cite{pytorch})}. 
However, in practice using auto-differentiation to compute the gradient is many times slower than the loss computation. 
\kledit{Instead we explicitly implement the gradient computation} $\frac{\partial\l L(\v L, \v X)}{\partial u_{t, s, v}}$ using the backward algorithm.  We define two backward variables $\beta^{(d)}_t$ and $\beta^{(n)}_{t,y}$ for the denominator and numerator, respectively:
\vspace{-0.1in}
\begin{equation}
\label{eq:beta_nom_denom}
    \begin{split}
    \beta^{(d)}_t &:=\displaystyle\sum_{\pi\in\l P_{t:T}}u(\pi)=\displaystyle\sum_{s=1}^{S}\displaystyle\sum_{v=1}^{V}u_{t,s,v}\beta^{(d)}_{t+s} \\[-5pt]
    \beta^{(n)}_{t,y} &:=\hspace{-15pt}\displaystyle\sum_{\substack{\pi\in\l P_{t:T} \\ \l B(\pi_{y:|\v L|})=\v L_{y:|\v L|}}}\hspace{-15pt}u(\pi)=\displaystyle\sum_{s=1}^{S}\displaystyle u_{t,s,l_y}\beta^{(n)}_{t+s,y+1}\\
    \end{split}
\end{equation}
The gradient $\frac{\partial L(\v L, \v X)}{\partial u_{t,s, v}}$ is then given by 
\vspace{-0.1in}
\begin{equation}
\label{eq:grad}
\begin{split}
& \frac{\partial \mathcal{L}(\v L, \v X)}{\partial u_{t, s, v}} = -\displaystyle\sum_{k\in \{k|l_{k}=v\}}\frac{\alpha^{(n)}_{t,k}\beta^{(n)}_{t+s,k}}{\alpha^{(n)}_{T,|\v L|}} + \frac{\alpha^{(d)}_{t} \beta^{(d)}_{t+s}}{\alpha^{(d)}_T}
\end{split}
\end{equation}
\ssedit{where $\{k|l_{k}=v\}$ \klnewedit{are the} 
indices in \ssedit{$\v L$} where label $v$ occurs.} 

\ssedit{
\subsection{Decoding}
\vspace{-0.1cm}
Decoding 
\kledit{consists} of solving $\pi^\star=\argmax_{\pi\in\mathcal{P}_{0:T}}w(\pi)$.
This optimization problem can be solved efficiently via the Viterbi algorithm with the recursive relationship:
\vspace{-0.05in}
\begin{equation}
\vspace{-0.05in}
  d(t) := \max_{\pi\in\mathcal{P}_{0:t}}w(\pi) = \max_{\substack{1\leq s\leq S\\ 1\leq v\leq V}}[w_{t-s,s,v}+d(t-s)]
\label{eq:viterbi}
\end{equation}
\ssedit{where} the last max \kledit{operation}
can be parallelized \kledit{on a GPU} such that the overall runtime\footnote{Number of times $\max(a, b)$ is called} of decoding is only $O(T\log(SV))$.
}

%\subsection{AWE and AGWE pre-training}
\subsection{\kledit{Pre-training via acoustic and acoustically grounded word embeddings}}
\label{ssec:agwe}
One important issue in whole-word models is that many words are infrequent or unseen in the training set.  In particular, the final weight layer, which corresponds to embeddings of the word labels, can be very poorly learned.  Recent work has shown that jointly pre-trained \klnewedit{acoustic word embeddings (AWEs)} and corresponding \klnewedit{acoustically grounded word embeddings (AGWEs)} of the written words~\cite{he+etal_iclr2017} can serve as a good parameter initialization for CTC-based A2W models~\cite{settle2019_a2w}, improving 
conversational speech recognition performance.  In this prior work, the AGWEs are parametric functions of character sequences, so that word embeddings can be produced for unseen or infrequent words.  We follow this idea and jointly pre-train our segmental 
\klnewedit{acoustic embedding function $f_{ac}$}
and the corresponding weight layer $\v A^{(2)}$ in Equation~\ref{eq:feat_func}. \klnewedit{This initialization is especially natural for whole-word segmental models, since the segments are explicitly intended to model words.}  Note that typical pre-trained written word embeddings (\kledit{such as} word2vec~\cite{mikolov2013efficient}, GloVe~\cite{pennington2014glove}, and contextual word embeddings~\cite{Peters:2018,devlin2018bert}) are not what is needed for the label embedding layer; we are interested in embeddings that represent the way a word sounds rather than what it means, \klnewedit{so acoustically grounded embeddings are the more natural choice}.

Our pre-training 
follows the multi-view AWE+AGWE training approach of~\cite{he+etal_iclr2017,settle2019_a2w}, \klnewedit{in which we} jointly train an acoustic ``view" embedding model %(the AWE model
($f$) 
%)
and a written \kledit{``view"} model ($g$) using a contrastive loss. 
The written view model
takes in a word label $v$, maps $v$ to a subword (e.g., character/phone) sequence \kledit{using a lexicon}, and uses this sequence to produce an 
\klnewedit{embedding vector} as output.  The resulting written word embedding model is "acoustically grounded" because it is learned jointly with the acoustic embedding model so as to represent the way the word sounds. 
\klnewedit{Specifically, we use}
an objective consisting of three contrastive triplet loss terms:
\vspace{-0.25cm}
\begin{flalign}
&\sum_{i=1}^{N}
    \left[
        m + d(f({\bf X}_i), g(v_i)) - \displaystyle \min_{v' \in \mathcal{V}_0'({\bf X}_i, v_i)} d(f({\bf X}_i), g(v'))
    \right]_{+}\notag\\
    +&\left[
        m + d(g(v_i), f({\bf X}_i)) - \displaystyle \min_{{\bf X}' \in \mathcal{X}_1'({\bf X}_i, v_i)} d(g(v_i), f({\bf X}'))
    \right]_{+}\notag\\
    +&\left[
        m + d(g(v_i), f({\bf X}_i)) - \displaystyle \min_{v' \in \mathcal{V}_2'({\bf X}_i, v_i)} d(g(v_i), g(v'))
    \right]_{+}
\label{eq:mv}
\end{flalign}
where ${\bf X}_i$ is a spoken word segment, $v_i$ is its word label, 
$m$ is a margin hyperparameter, $d$ denotes cosine distance $d(a,b) = 1-\frac{a\cdot b}{\Vert a \Vert\Vert b \Vert}$, and $N$ is the number of training pairs %$({\bf X}, y)$.
$({\bf X}, v)$. We conduct semi-hard~\cite{schroff2015facenet}
negative sampling w.r.t.~each %
pair:
\begin{flalign*}
    \mathcal{V}_0'({\bf X}, v) &:= \{
        v' \vert d(f({\bf X}), g(v')) > d(f({\bf X}), g(v)), v' \in \mathcal{V} / v
    \}\\
    \mathcal{X}_1'({\bf X}, v) &:= \{
        {\bf X}' \vert d(g(v), f({\bf X}')) > d(g(v), f({\bf X})), v' \in \mathcal{V} / v 
    \}\\
    \mathcal{V}_2'({\bf X}, v) &:= \{
        v' \vert d(g(v), g(v')) > d(g(v), f({\bf X})), v' \in \mathcal{V} / v
    \}
\end{flalign*}
where $\mathcal{V}$ is the training vocabulary and $v'$ is the word label of ${\bf X}'$.
For efficiency, this \klnewedit{negative sampling} is performed over the mini-batch such that $N$ is the batch size and $\mathcal{V}$ consists of words in the mini-batch. Additionally, rather than the single most offending semi-hard negative we use $M$ and each contrastive loss term inside the sum in Equation~\ref{eq:mv} is an average over these $M$ negatives.
The contrastive loss aims to map spoken word segments corresponding to the same word label close together and close to their learned label embeddings, while ensuring that segments corresponding to different word labels are mapped farther apart (and nearer to their respective label embeddings). Our pre-training approach is the same as that of~\cite{he+etal_iclr2017,settle2019_a2w} except for the addition of semi-hard negative sampling (replacing hard negative sampling in~\cite{settle2019_a2w}), the inclusion of a third contrastive term ($obj_1$ of~\cite{he+etal_iclr2017}), \kledit{and an extra convolutional layer and pooling in the AWE encoder (see Section~\ref{sec:exp}).} 
The first two changes %were found to
increase 
\klnewedit{word discrimination task} performance %
in prior work \kledit{on AWEs}~\cite{settle2016discriminative,Kamper_16a,he+etal_iclr2017}, and the third change improves \kledit{efficiency of the segmental model.} 

 The pre-trained AWE/AGWE models are tuned using a cross-view word discrimination task as in~\cite{he+etal_iclr2017,settle2019_a2w}, applied to word segments from the development set and word labels from the 
 vocabulary. The task is to determine whether a given acoustic word segment and word label match.
 We compute the embeddings of the acoustic segment and character sequence by forwarding them through $f$ and $g$, respectively, and then compute their cosine distance. If this distance is below a threshold, then the pair is labeled a match. The quality of the embeddings is measured by the average precision (AP) over all thresholds over the dev set. 

 Similarly to~\cite{settle2019_a2w}, in addition to initializing with the pre-trained AGWEs, we also consider $L2$ regularization toward the pre-trained AGWEs. 
\kledit{We} add a term to the recognizer loss (Equation~\ref{eq:log_loss}) corresponding to the distance between the rows $\v a^{(2)}_v$ of $\v A^{(2)}$ and the pre-trained AGWEs $g(v)$: %
\vspace{-0.2cm}
\begin{equation}
\label{eq:reg_agwe}
\mathcal{L}_{reg}(\v L, \v X)=(1-\lambda)\l L_{seg}(\v L, \v X)+\lambda\displaystyle\sum_{v\in\v L}\|\v a^{(2)}_v - g(v)\|^2
\end{equation}
where $\lambda$ is a hyperparameter. 
\vspace{-0.3cm}

\section{Experiments}
\label{sec:exp}
\vspace{-0.15cm}

\klnewedit{We conduct experiments on the standard Switchboard-300h dataset and data division~\cite{godfrey1992switchboard}. We use 40-dimensional log-Mel spectra +$\Delta$+$\Delta\Delta$s, extracted with Kaldi~\cite{povey2011kaldi}, as input features.} \kledit{Every two} successive frames are stacked and  alternate frames dropped, resulting in 240-dimensional features.
We explore $5K$, $10K$, and $20K$ vocabularies based on word occurrence thresholds of $18$, $6$, and $2$, respectively. \klnewedit{Hyperparameters are chosen based on prior related work (e.g.,~\cite{settle2019_a2w}) and light tuning on the Switchboard development set.}
The backbone network for the segmental model is a 6-layer bidirectional long short-term memory network (BiLSTM) with 512 hidden units per direction \kledit{per layer} with dropout added between layers (0.25 \klnewedit{except when otherwise specified below}).
\kledit{To speed up training,} we add a convolutional layer with kernel size 5 followed by average pooling with stride 4 on top of the BiLSTM. The maximum segment length is set to 32,  
\kledit{corresponding to a maximum word duration 
of} $\sim2.4s$. 
To further speed up training, we reduce the maximum segment size per 
batch\bsedit{ (batch size: 16)} to $\min\{2*\max\{\frac{\text{input length}}{\text{\# words}}\}, 32\}$.
\bsedit{The model is trained  with the Adam optimizer~\cite{kingma2014adam} with an initial learning rate of 0.001, which is decreased by a factor of 2 when the dev WER stops decreasing.  No language model is used for decoding.} 
\ssedit{
\kledit{As a baseline, we also train an A2W CTC model using the same 
structure (6-layer BiLSTM + convolutional + pooling).}
}

\begin{figure}[t]
\centering
\includegraphics[width=\linewidth]{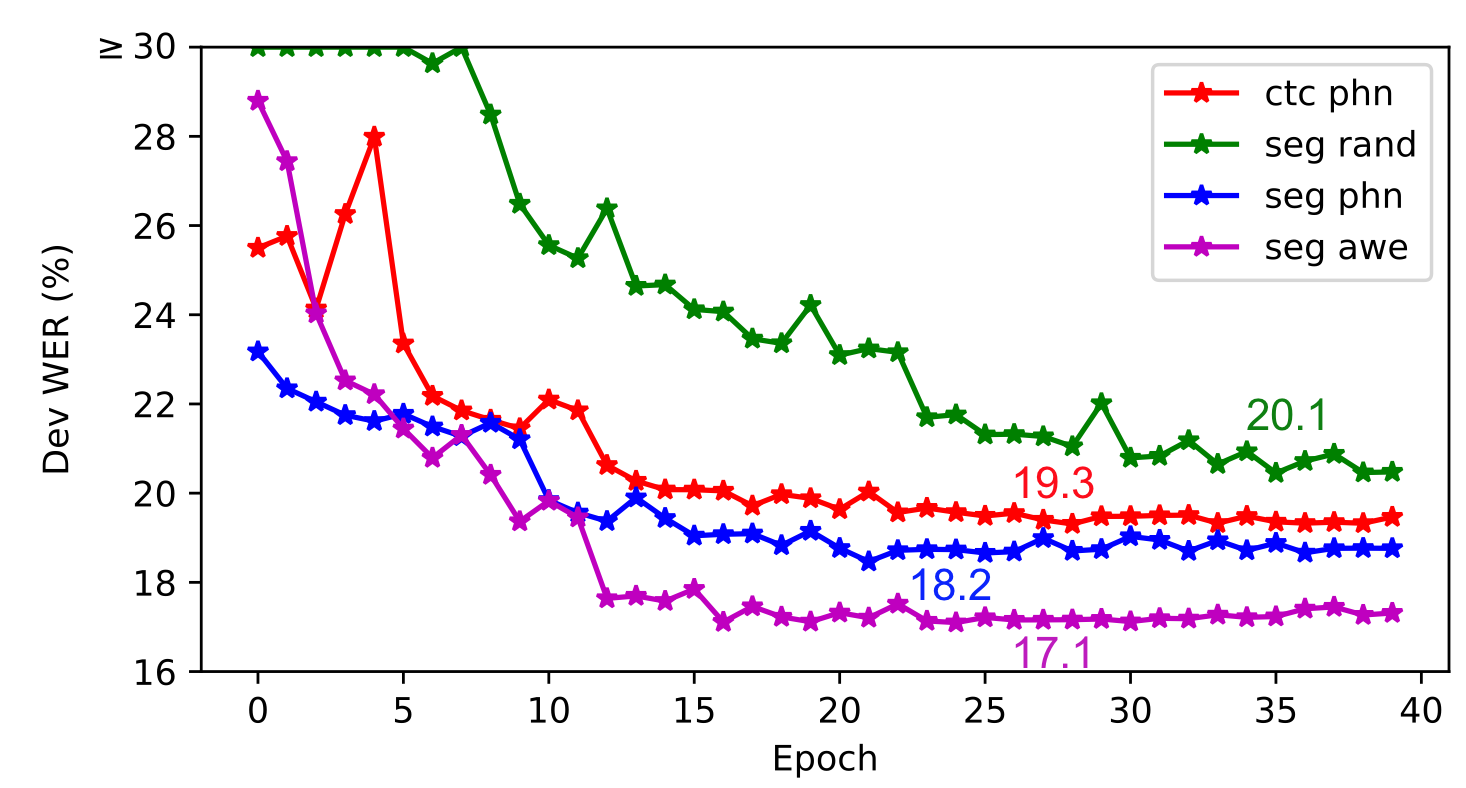}
\vspace{-0.7cm}\caption{\label{fig:ctc_seg_dev}Dev WER vs.~epoch for A2W CTC and segmental models with different initialization. Numbers: lowest dev WER.}
\vspace{-0.3cm}
\end{figure}

\vspace{-0.3cm}
\subsection{\label{ssec:exp_phn}Phone CTC pre-training}
\vspace{-0.1cm}
\kledit{As an initial experiment with the 5K vocabulary, we} initialize the backbone BiLSTM network by pre-training with a phone CTC objective, \kledit{as in prior work on CTC-based A2W models}~\cite{audhkhasi2017_a2w,audhkhasi2018_a2w,yu2018_amt,settle2019_a2w}. The phone error rate (PER) of this 
phone CTC model is $11.0\%$. 
On top of the pre-trained BiLSTM, the convolutional layer and word embedding parameters 
($\v A^{(1)}, \v A^{(2)}, \v b^{(1)}, \v b^{(2)}$) in our segmental model are randomly initialized.  Compared to random initialization, \kledit{phone CTC pre-training reduces WER by} $1.9\%$ \ssedit{(Figure~\ref{fig:ctc_seg_dev})}, which is consistent with 
\kledit{prior work}~\cite{audhkhasi2017_a2w}.
\kledit{When both our segmental model and the baseline A2W CTC model are pre-trained with phone CTC,
our model 
achieves $\sim1\%$ lower word error rate (WER)} (Figure~\ref{fig:ctc_seg_dev}).

The pooling operation $G$ in Equation~\ref{eq:awe_func} is tuned among mean pooling (18.5\%), attention pooling (19.0\%) and concatenation (18.2\%). 
\klnewedit{ The best performance is obtained with concatenation, 
which also increases the feature dimensionality of $\v A^{(1)}$,
while consuming a factor of 
$S/2$ less memory when computing segment features.} 

\begin{table}[t]
  \caption{\label{tab:exp_agwe_init}\kledit{Comparison of initialization with phone CTC vs.~\kledit{AWE/AGWE}, in terms of SWB dev WER (\%).  ``AWE init" refers to initialization of the parameters of \klnewedit{$f_{ac}$}.
  ``AGWE init" refers to 
  \bsnewedit{initialization of 
  ${\bf A}^{(2)}$.} }
  \vspace{-0.3cm}
  }
  \centering
  \begin{tabular}{lccc}\\\toprule
  \multirow{2}{*}{System} & \multicolumn{3}{c}{Vocab size} \\ & 5K & 10K & 20K \\\midrule
    A2W CTC with phone CTC init & 19.3 & 18.0 & 17.7 \\
    \midrule
    A2W Segmental with phone CTC init  & 18.2 & 17.9 & 18.0 \\
    \hspace{0.25cm} + AGWE init & 18.4 & 18.0 & 18.0 \\
    A2W Segmental with AWE init & 17.1 & 16.0 & 16.4 \\
    \hspace{0.25cm} + AGWE init & 17.1 & 15.8 & 16.5 \\
    \hspace{0.25cm} + AGWE $L2$ reg & \textbf{17.0} & \textbf{15.5} & \textbf{15.6}\\
    \bottomrule
  \end{tabular}
    \vspace{-0.5cm}
\end{table}

\vspace{-.2in}
\subsection{Vocabulary size}
\label{ssec:vocab-size}
\vspace{-.05in}
 We find that, unlike our A2W CTC models and those of prior work~\cite{settle2019_a2w,yu2018_amt}, the segmental models do not \klnewedit{necessarily} improve
 with larger vocabulary.
One possible reason is that word representations in segmental models, especially for rare words, are harder to learn as they must be robust to variations in segment duration \kledit{and content}.
Segmental models \ssedit{may} require \klnewedit{more data 
when} many rare words are included.
\bsedit{% 
We find that it is important to set a larger dropout value as the vocabulary size increases.}   
\kledit{The best dropout values for $5K$, $10K$ and $20K$ are $0.25$, $0.35$, and $0.45$, respectively, with results in Table~\ref{tab:exp_agwe_init}.}

\vspace{-.05in}
\subsection{\klnewedit{AWE + AGWE} pre-training}
\label{ssec:agwe-expts}
\vspace{-.05in}

 We now investigate whether \klnewedit{pre-training with AWE and AGWEs} can provide a better starting point for the segmental model.
We jointly train AWE and AGWE models on the Switchboard-300h training set with the multi-view training approach described in Section~\ref{ssec:agwe}. \klnewedit{Early stopping and hyperparameter tuning are done based on the cross-view average precision (AP) on the same development set as in ASR training.  In the contrastive training objective, we use $m=0.45$, $M=64$ reduced by $1$ per batch until $M=6$, and a variable batch size with up to $20,000$ frames per batch.}  The acoustic view model $f$ has the same structure as \klnewedit{$f_{ac}$ in the segmental model,
and the written view model $g$ is composed of an input embedding layer mapping 
$37$ input characters} to $32$-dimensional embeddings followed by a 1-layer BiLSTM with 256 hidden units per direction. We optimize with the Adam optimizer~\cite{kingma2014adam}, with an initial learning rate of 0.0005, which is reduced by a factor of 10 when the development set cross-view AP does not improve for $3000$ steps.  Training is stopped when the learning rate drops below $10^{-9}$. 
\ssedit{After multi-view training, the acoustic view $f$ (our AWE function) and the written view $g$ (our AGWE function) are used to initialize our segmental feature function 
\klnewedit{$f_{ac}$} and $\v A^{(2)}$, respectively, in Equation~\ref{eq:feat_func}.} 

\begin{figure}[t]
\centering
\includegraphics[width=\linewidth]{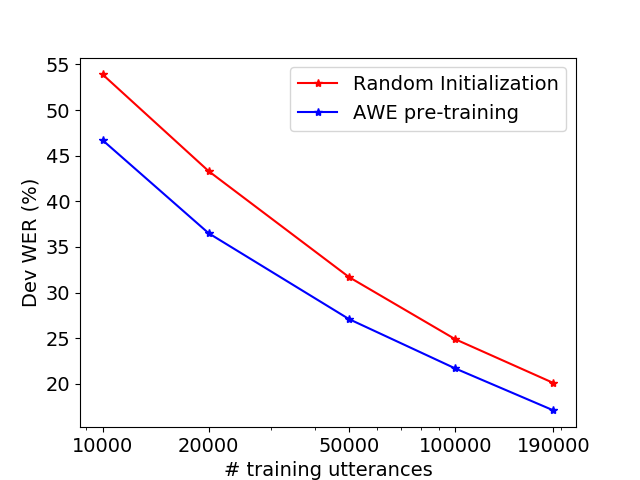}
\vspace{-0.6cm}
\caption{\label{fig:wer_utt}\klnewedit{Segmental model dev WER, using random initialization vs.~AWE pre-training (with AGWE initialization) with various ASR training set sizes, using $5K$-word vocabulary.}}
\vspace{-0.3cm}
\end{figure}

\begin{figure}[t]
% \vspace{-0.2cm}
\centering
\includegraphics[width=\linewidth]{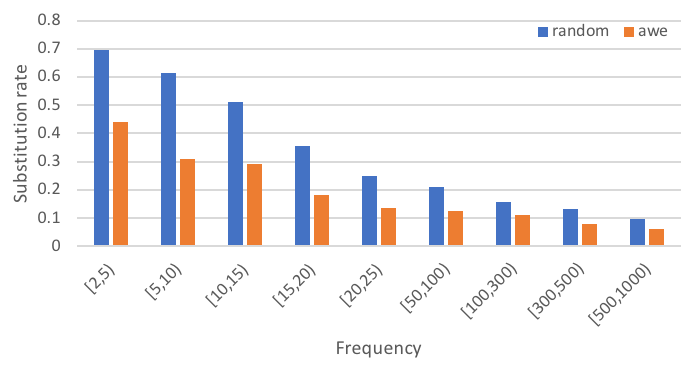}
\vspace{-0.6cm}
\caption{Word substitution rates when using AWE pre-training vs.~random initialization, for words with different training set frequencies. The vocabulary size is $20K$.}
\label{fig:error_frequency}
\vspace{-0.5cm}
\end{figure}

Table~\ref{tab:exp_agwe_init} compares an A2W CTC model with segmental models using different initializations, \klnewedit{evaluated on the SWB development set}.
Initialization of $f_{ac}$ with the pre-trained AWE model 
reduces WER by $1$--$2\%$ over phone CTC initialization.
\kledit{Initialization of $\v A^{(2)}$ with pre-trained AGWE models alone does not help, but initializing with AWE and AGWE while regularizing toward the pre-trained AGWEs (see Section~\ref{ssec:agwe}) is helpful, especially for larger vocabularies.  This observation is consistent with our expectation:  Since the AGWEs are composed from character sequences, they are less impacted by vocabulary size, helping with recognition of rare words.} 
\klnewedit{We also note that the} \kledit{optimal $\lambda$ in Equation~\ref{eq:reg_agwe} tends to be larger as the vocabulary size increases, reinforcing the need for more regularization when there are many rare words.}
\klnewedit{This intuition about rare words also suggests that AWE pre-training should be more helpful for smaller training sets and for rarer words.
\bsnewedit{Figures~\ref{fig:wer_utt} and ~\ref{fig:error_frequency} demonstrate this expected result.}
} 

\vspace{-.05in}
\subsection{\klnewedit{Final evaluation results}}
\label{ssec:final-results}
\vspace{-.05in}

 Table~\ref{tab:exp_eval2000} shows the final results on the Switchboard (SWB) and CallHome (CH) test sets, compared to prior work with A2W models.\footnote{We do not compare with prior segmental models~\cite{tang2014_losses,lu2016_srnn} \klnewedit{since, due to their larger memory footprint}, we are unable to train them for A2W recognition with a \bsnewedit{similar} network architecture using a typical GPU (e.g., 12GB memory).} For the smaller vocabularies ($5K$, $10K$), the segmental model improves WER over CTC by around $1\%$ (absolute).  \bsedit{Training with SpecAugment~\cite{park2019specaugment} produces an additional gain of $\sim 1\%$ across all vocabulary sizes.} 
  \klnewedit{As noted before, the $10K$-word model outperforms the $20K$-word one.   % 
 In addition to the issue of rare words, the longer training time with a $20K$-word vocabulary prevents us from tuning hyperparamters as much as for the $10K/5K$ models.}  \klnewedit{Overall our best model improves over all previous A2W models of which we are aware, although our model is smaller than the previous best-performing model~\cite{yu2018_amt}.}

 Despite the improved performance of our A2W models, there is still a gap between our (and all) A2W models and systems based on subword units, where the best performance on this task of which we are aware is 6.3\% on SWB and 13.3\% on CH using a sub-word based sequence-to-sequence transformer model~\cite{park2019specaugment}.  Some prior work suggests that the gap between between A2W and subword-based models can be significantly narrowed when using larger training sets, and our future work includes investigations with larger data.  At the same time, A2W models retain the benefit of being very simple, truly end-to-end models that avoid using a decoder or potentially a language model.

\begin{table}[t]
\vspace{-0.4cm}
    \caption{\label{tab:exp_eval2000}WER (\%) results on SWB/CH evaluation sets.}
    \vspace{-0.3cm}
  \small
  \setlength{\tabcolsep}{3pt}% \arraystretch{2}
  \centering
  \begin{tabular}{lccc}\\\toprule
    \multirow{2}{*}{System} & \multicolumn{3}{c}{Vocab} \\
     & 4K/5K & 10K & 20K \\\midrule
    \footnotesize{Seg, AWE+AGWE init} & 14.0/24.9 & 12.8/23.5 & 12.5/24.5 \\ 
    \footnotesize{\hspace{0.25cm}+SpecAugment} & \textbf{12.8/22.9} & \textbf{10.9/20.3} & 12.0/21.9 \\ \midrule
    \footnotesize{CTC, phone init \cite{settle2019_a2w}} &  16.4/25.7 & 14.8/24.9 & 14.7/24.3 \\
    \footnotesize{CTC, AWE+AGWE init \cite{settle2019_a2w}} & 15.6/25.3 & 14.2/24.2 & 13.8/24.0 \\
    \footnotesize{\hspace{0.25cm}+reg\cite{settle2019_a2w}} & 15.5/25.4 & 14.0/24.5 & 13.7/23.8 \\
    \footnotesize{CTC, AWE+AGWE rescore \cite{settle2019_a2w}} & 15.0/25.3 & 14.4/24.5 & 14.2/24.7 \\
    \footnotesize{S2S \cite{palaskar2018_s2s}} &  - & 22.4/36.1 & 22.4/36.2 \\
    \footnotesize{Curriculum \cite{yu2018_amt}}  & - & - & 13.4/24.2 \\
    \footnotesize{\hspace{0.25cm}+Joint CTC/CE\cite{yu2018_amt}}  & - & - &  13.0/23.4 \\
    \footnotesize{\hspace{0.25cm}+Speed Perturbation\cite{yu2018_amt}}  & - & - &  \textbf{11.4/20.8} \\
    \bottomrule
  \end{tabular}
  \vspace{-0.4cm}
\end{table}
%   \vspace{-0.5cm}

\vspace{-0.3cm}
\subsection{Decoding and training speed}

\begin{table}[t]
  \caption{\klnewedit{Training/decoding latency (in ms) of our segmental model, CTC, and other segmental models} \kledit{with a 10K-word vocabulary}\label{tab:latency}. Numbers in () in the training and decoding columns are time consumed by the loss forward/backward algorithm and the actual decoding algorithm, respectively. The batch size for training and decoding are 16 and 1 respectively. 
   \small{CPU: Intel(R) Core(TM) i7-3930K CPU @ 3.20GHz.} GPU: \small{Titan X.} }
    \vspace{-0.2cm}
    \small
  \setlength{\tabcolsep}{5pt}%
  \begin{tabular}{lrrr}\\\toprule
    Model & \multicolumn{1}{c}{training} & decoding, GPU & decoding, CPU \\  \midrule
  CTC & 368.2 (40.3) & 61.0 \hspace{4pt}(0.6) & 634.5 \hspace{4pt}(0.6) \\
  Seg, ours & 509.6 (69.1) & 92.9 (32.7) & 1009.8 (60.1)\\
  Seg, SRNN & 1739.7 (72.8) & 120.9 (35.4) & 1232.5 (57.4)\\
  Seg, FC & 1361.5 (71.5) & 109.7 (33.2) & 1031.4 (60.4)\\
  
    \bottomrule
  \end{tabular}
  \vspace{-0.3cm}
\end{table}

\begin{figure}[t]
\centering
\includegraphics[width=\linewidth]{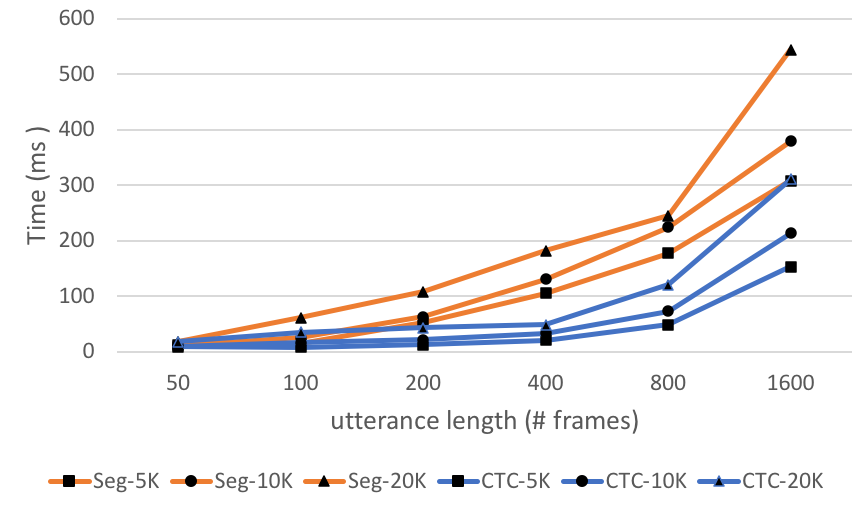}
\vspace{-0.8cm}
\caption{\label{fig:latency}Comparison of loss computation time (forward/backward) of CTC and our segmental model.}
\vspace{-0.5cm}
\end{figure}

Our implementation is based on
PyTorch, with the forward/backward computation for the segmental
loss implemented in CUDA C. Training on the 300-hour Switchboard training set takes about 2 days on one
Titan X GPU. Figure \ref{fig:latency} shows the latency of the loss forward/backward computation compared with CTC. For CTC we use the Warp-CTC implementation in ESPnet~\cite{espnet}. The segmental loss forward/backward computation is roughly $0.5$x slower than for CTC, which is mainly due to the \klnewedit{denominator computations ($\beta_t^{(d)}$ and $\alpha_t^{(d)}$ in Equations \ref{eq:alpha_nom_denom}, \ref{eq:beta_nom_denom})}. We also measure training time per batch and decoding time per utterance averaged over 100 random batches from the dev set (see Table \ref{tab:latency}). 

\klnewedit{To demonstrate the efficiency improvement enabled by our segmental feature function, we also compare to (our implementation of) \klnewedit{a whole-word SRNN~\cite{lu2016_srnn} and a whole-word frame classifier (FC)-based segmental model}~\cite{tang2017_e2e} with the same backbone network.} 
\klnewedit{The SRNN/FC-based models differ from our segmental model in terms of the feature functions, but we use the same 
implementation of the forward/backward/Viterbi algorithms for all three models.  We note that, in order to make the speed comparison with an FC-based model possible, we simplified it somewhat 
so as to fit it into GPU memory.\footnote{Specifically, we removed the frame average feature function (Equation 6 in~\cite{tang2017_e2e}) and added an extra linear layer on top of the BiLSTM to reduce dimensionality of $h_s$ and $h_t$ to 1 (first equation in Section \MakeUppercase{\romannumeral 3}.B in~\cite{tang2017_e2e}).}  These two models are implemented only for this speed test; we fail to train them to completion because, even with our simplifications, we still run out of memory for \klnewedit{some 
training utterances.}}
Overall our segmental model is roughly $0.5$x slower to train than CTC 
but more than twice faster than \klnewedit{the SRNN/FC-based models.}
In practice, the loss computation is a small fraction of the total training time \klnewedit{(see
Table \ref{tab:latency})}. 
Computing the segment score function is the main factor that accounts for the larger training latency compared to CTC. 

Similarly, our segmental model is \klnewedit{also} $0.5$x slower than CTC in decoding. Compared to CTC, the actual decoding algorithm accounts for a larger proportion of the total latency. The CTC greedy decoding can be parallelized more ($O(\log(V))$) than Viterbi decoding ($O(T\log(SV))$).
However, using a GPU results in larger speed gains for Viterbi decoding.

\vspace{-0.25cm}
\section{\klnewedit{Conclusion}}
\vspace{-0.2cm}
We have introduced an end-to-end whole-word segmental model, which to our knowledge is the first to perform large-vocabulary speech recognition competitively and efficiently. Our model uses a simple segment score function based on a dot product between written word embeddings and acoustic \klnewedit{segment embeddings, which both improves efficiency and enables us to pre-train the model with jointly trained acoustic and written word embeddings.}  
We find that the proposed model outperforms \klnewedit{previous A2W approaches, and is much more efficient than previous segmental models.  The key aspects that are important to the performance improvements are pre-training the acoustic segment representation with acoustic word embeddings and regularizing the label embeddings toward pre-trained acoustically grounded word embeddings}. 
Given the good performance of segmental models especially when the label set is relatively small, it will also be interesting to apply the approach to recognition based on subwords like byte pair encodings~\cite{sennrich2016neural} \klnewedit{or other more acoustically motivated subwords~\cite{xu2019improving}, and to study the applicability of our models to a wider range of training set sizes and domains}. 
\vspace{-0.25cm}

\section{Acknowledgements}
\vspace{-0.25cm}
We are grateful to Hao Tang for helpful suggestions and discussions. This research was funded by NSF award IIS-1816627, and by an AWS Machine Learning Research Award.

\bibliographystyle{IEEEbib}
% \bibliography{strings,refs}
\bibliography{mybib}

\begin{thebibliography}{10}

\bibitem{soltau_a2w}
H.~Soltau, H.~Liao, and H.~Sak,
\newblock ``Neural speech recognizer: Acoustic-to-word {LSTM} model for large
  vocabulary speech recognition,''
\newblock in {\em Proc.~{I}nterspeech}, 2016.

\bibitem{audhkhasi2017_a2w}
K.~Audhkhasi, B.~Ramabhadran, G.~Saon, M.~Picheny, and D.~Nahamoo,
\newblock ``Direct acoustics-to-word models for {English} conversational speech
  recognition,''
\newblock in {\em Proc.~{I}nterspeech}, 2017.

\bibitem{audhkhasi2018_a2w}
Kartik Audhkhasi, Brian Kingsbury, Bhuvana Ramabhadran, George Saon, and
  Michael Picheny,
\newblock ``Building competitive direct acoustics-to-word models for {English}
  conversational speech recognition,''
\newblock in {\em Proc.~{I}nterspeech}, 2018.

\bibitem{yu2018_amt}
Chengzhu Yu, Chunlei Zhang, Chao Weng, Jia Cui, and Dong Yu,
\newblock ``A multistage training framework for acoustic-to-word model,''
\newblock in {\em Proc.~{I}nterspeech}, 2018.

\bibitem{settle2019_a2w}
Shane Settle, Kartik Audhkhasi, Karen Livescu, and Michael Picheny,
\newblock ``Acoustically grounded word embeddings for improved
  acoustics-to-word speech recognition,''
\newblock in {\em Proc. ICASSP}, 2019.

\bibitem{li2018advancing}
Jinyu Li, Guoli Ye, Amit Das, Rui Zhao, and Yifan Gong,
\newblock ``Advancing acoustic-to-word {CTC} model,''
\newblock in {\em Proc. ICASSP}, 2018.

\bibitem{gaur2019acoustic}
Yashesh Gaur, Jinyu Li, Zhong Meng, and Yifan Gong,
\newblock ``Acoustic-to-phrase models for speech recognition,''
\newblock in {\em Proc.~{I}nterspeech}, 2019.

\bibitem{graves2006connectionist}
Alex Graves, Santiago Fern{\'a}ndez, Faustino Gomez, and J{\"u}rgen
  Schmidhuber,
\newblock ``Connectionist temporal classification: labelling unsegmented
  sequence data with recurrent neural networks,''
\newblock in {\em Proc. Int. Conf. on Machine Learning ({ICML})}, 2006.

\bibitem{collobert2019wordlevel}
Ronan Collobert, Awni Hannun, and Gabriel Synnaeve,
\newblock ``Word-level speech recognition with a dynamic lexicon,''
\newblock {\em arXiv preprint arXiv:1906.04323}, 2019.

\bibitem{palaskar2018_s2s}
Shruti Palaskar and Florian Metze,
\newblock ``Acoustic-to-word recognition with sequence-to-sequence models,''
\newblock in {\em Proc.~{IEEE} Workshop on Spoken Language Technology ({SLT})},
  2018.

\bibitem{ostendorf1996hmm}
Mari Ostendorf, Vassilios Digalakis, and Owen Kimball,
\newblock ``From {HMM}'s to segment models: a unified view of stochastic
  modeling for speech recognition,''
\newblock {\em IEEE Trans. Speech and Audio Processing}, vol. 4, pp. 360--378,
  1996.

\bibitem{glass2003probabilistic}
James~R Glass,
\newblock ``A probabilistic framework for segment-based speech recognition,''
\newblock {\em Computer Speech \& Language}, vol. 17, no. 2-3, pp. 137--152,
  2003.

\bibitem{zweig2009segmental}
Geoffrey Zweig and Patrick Nguyen,
\newblock ``A segmental {CRF} approach to large vocabulary continuous speech
  recognition,''
\newblock in {\em Proc.~{IEEE} Workshop on Automatic Speech Recognition and
  Understanding ({ASRU})}, 2009.

\bibitem{zweig2012_seg}
Geoffrey Zweig,
\newblock ``Classification and recognition with direct segment models,''
\newblock {\em Proc. ICASSP}, pp. 4161--4164, 2012.

\bibitem{he2012_sc}
Yanzhang He and Eric Fosler-Lussier,
\newblock ``Efficient segmental conditional random fields for phone
  recognition,''
\newblock in {\em Proc.~{I}nterspeech}, 2012.

\bibitem{hamid2013_dsnn}
O.~Abdel-Hamid, L.~Deng, D.~Yu, and H.~Jiang,
\newblock ``Deep segmental neural networks for speech recognition,''
\newblock in {\em Proc.~{I}nterspeech}, 2013.

\bibitem{tang2014_losses}
Hao Tang, Kevin Gimpel, and Karen Livescu,
\newblock ``A comparison of training approaches for discriminative segmental
  models,''
\newblock in {\em Proc.~{I}nterspeech}, 2014.

\bibitem{lu2016_srnn}
Liang Lu, Lingpeng Kong, Chris Dyer, Noah~A. Smith, and Steve Renals,
\newblock ``Segmental recurrent neural networks for end-to-end speech
  recognition,''
\newblock in {\em Proc.~{I}nterspeech}, 2016.

\bibitem{maas2012word}
Andrew~L Maas, Stephen~D Miller, Tyler~M O’neil, Andrew~Y Ng, and Patrick
  Nguyen,
\newblock ``Word-level acoustic modeling with convolutional vector
  regression,''
\newblock in {\em Proc. ICML Workshop on Representation Learning}, 2012.

\bibitem{bengio2014word}
Samy Bengio and Georg Heigold,
\newblock ``Word embeddings for speech recognition,''
\newblock in {\em Proc. ICASSP}, 2014.

\bibitem{settle2017query}
Shane Settle, Keith Levin, Herman Kamper, and Karen Livescu,
\newblock ``Query-by-example search with discriminative neural acoustic word
  embeddings,''
\newblock in {\em Proc.~{I}nterspeech}, 2017.

\bibitem{audhkhasi2017end}
Kartik Audhkhasi, Andrew Rosenberg, Abhinav Sethy, Bhuvana Ramabhadran, and
  Brian Kingsbury,
\newblock ``End-to-end {ASR}-free keyword search from speech,''
\newblock {\em IEEE Journal of Selected Topics in Signal Processing}, vol. 11,
  no. 8, pp. 1351--–1359, 2017.

\bibitem{tang2015_cascade}
Hao Tang, Weiran Wang, Kevin Gimpel, and Karen Livescu,
\newblock ``Discriminative segmental cascades for feature-rich phone
  recognition,''
\newblock in {\em Proc.~{IEEE} Workshop on Automatic Speech Recognition and
  Understanding ({ASRU})}, 2015.

\bibitem{tang2017_e2e}
Hao Tang, Liang Lu, Kevin Gimpel, Chris Dyer, and Ann~S. Smith,
\newblock ``End-to-end neural segmental models for speech recognition,''
\newblock {\em IEEE Journal of Selected Topics in Signal Processing}, 2017.

\bibitem{kong2016_srnn}
Lingpeng Kong, Chris Dyer, and Noah~A. Smith,
\newblock ``Segmental recurrent neural networks,''
\newblock in {\em Proc. Int. Conf. on Learning Representations (ICLR)}, 2016.

\bibitem{he+etal_iclr2017}
W.~He, W.~Wang, and K.~Livescu,
\newblock ``Multi-view recurrent neural acoustic word embeddings,''
\newblock in {\em Proc. Int. Conf. on Learning Representations (ICLR)}, 2017.

\bibitem{zweig2011speech}
Geoffrey Zweig et~al.,
\newblock ``Speech recognition with segmental conditional random fields: A
  summary of the {JHU CLSP} 2010 summer workshop,''
\newblock in {\em Proc. ICASSP}, 2011.

\bibitem{pytorch}
Adam Paszke et~al.,
\newblock ``{PyTorch}: An imperative style, high-performance deep learning
  library,''
\newblock in {\em Advances in Neural Information Processing Systems (NIPS)},
  2019.

\bibitem{mikolov2013efficient}
Tomas Mikolov, Kai Chen, Gregory~S. Corrado, and Jeffrey Dean,
\newblock ``Efficient estimation of word representations in vector space,''
\newblock {\em CoRR}, vol. abs/1301.3781, 2013.

\bibitem{pennington2014glove}
J.~Pennington, R.~Socher, and C.~Manning,
\newblock ``Glo{V}e: Global vectors for word representation,''
\newblock in {\em Proc.~Conf. on Empirical Methods in Natural Language
  Processing ({EMNLP})}, 2014.

\bibitem{Peters:2018}
Matthew~E. Peters, Mark Neumann, Mohit Iyyer, Matt Gardner, Christopher Clark,
  Kenton Lee, and Luke Zettlemoyer,
\newblock ``Deep contextualized word representations,''
\newblock in {\em Proc. NAACL}, 2018.

\bibitem{devlin2018bert}
Jacob Devlin, Ming-Wei Chang, Kenton Lee, and Kristina Toutanova,
\newblock ``{BERT}: Pre-training of deep bidirectional transformers for
  language understanding,''
\newblock in {\em Proc. NAACL}, 2019.

\bibitem{schroff2015facenet}
F.~Schroff, D.~Kalenichenko, and J.~Philbin,
\newblock ``{FaceNet}: A unified embedding for face recognition and
  clustering,''
\newblock in {\em Proc. IEEE Conf. Computer Vision and Pattern Recognition
  (CVPR)}, 2015.

\bibitem{settle2016discriminative}
S.~Settle and K.~Livescu,
\newblock ``Discriminative acoustic word embeddings: Recurrent neural
  network-based approaches,''
\newblock in {\em Proc.~{IEEE} Workshop on Spoken Language Technology ({SLT})},
  2016.

\bibitem{Kamper_16a}
H.~Kamper, W.~Wang, and K.~Livescu,
\newblock ``Deep convolutional acoustic word embeddings using word-pair side
  information,''
\newblock in {\em Proc. ICASSP}, 2016.

\bibitem{godfrey1992switchboard}
John~J Godfrey, Edward~C Holliman, and Jane McDaniel,
\newblock ``{SWITCHBOARD}: Telephone speech corpus for research and
  development,''
\newblock in {\em Proc. ICASSP}, 1992.

\bibitem{povey2011kaldi}
Daniel Povey, Arnab Ghoshal, Gilles Boulianne, Lukas Burget, Ondrej Glembek,
  Nagendra Goel, Mirko Hannemann, Petr Motlicek, Yanmin Qian, Petr Schwarz,
  et~al.,
\newblock ``The {Kaldi} speech recognition toolkit,''
\newblock in {\em Proc.~{IEEE} Workshop on Automatic Speech Recognition and
  Understanding ({ASRU})}, 2011.

\bibitem{kingma2014adam}
Diederik Kingma and Jimmy Ba,
\newblock ``Adam: A method for stochastic optimization,''
\newblock in {\em Proc. Int. Conf. on Learning Representations (ICLR)}, 2014.

\bibitem{park2019specaugment}
Daniel~S. Park, William Chan, Yu~Zhang, Chung-Cheng Chiu, Barret Zoph, Ekin~D.
  Cubuk, and Quoc~V. Le,
\newblock ``Spec{A}ugment: A simple data augmentation method for automatic
  speech recognition,''
\newblock in {\em Proc.~{I}nterspeech}, 2019.

\bibitem{espnet}
Shinji Watanabe et~al.,
\newblock ``{ESP}net: End-to-end speech processing toolkit,''
\newblock in {\em Proc.~{I}nterspeech}, 2018.

\bibitem{sennrich2016neural}
Rico Sennrich, Barry Haddow, and Alexandra Birch,
\newblock ``Neural machine translation of rare words with subword units,''
\newblock in {\em Proc.~Association for Computational Linguistics}, 2016.

\bibitem{xu2019improving}
Hainan Xu, Shuoyang Ding, and Shinji Watanabe,
\newblock ``Improving end-to-end speech recognition with pronunciation-assisted
  sub-word modeling,''
\newblock in {\em Proc. ICASSP}, 2019.

\end{thebibliography}

\end{document}